# Dimensionality-Driven Anomalous Metallic State with Zero-field Nonreciprocal Transport in Layered Ising Superconductors

Yanwei Cui,[1,*] Zenglin Liu,[1,*] Qin Liu,[1,*] Junlin Xiong,[1,*], Yongqin Xie,[1] Yudi Dai,[1] Ji Zhou,[1] Lizheng Wang,[1] Hanyan Fang,[1] Haiwen Liu,[3] Shi-Jun Liang,[1] Bin Cheng[1,2,†] and Feng Miao[1,‡]

[1]National Laboratory of Solid State Microstructures, Institute of Brain-Inspired Intelligence, School of Physics, Collaborative Innovation Center of Advanced Microstructures, Nanjing University, Nanjing 210093, China.
[2]Institute of Interdisciplinary Physical Sciences, School of Science, Nanjing University of Science and Technology, Nanjing 210094, China.
[3]Center for Advanced Quantum Studies, Department of Physics, Beijing Normal University, Beijing 100875, China.

*Abstract*—The anomalous metal state (AMS), observed in "failed" superconductors, provides insights into superconductivity and quantum criticality, with studies revealing unconventional quantum phases like the Bose metal. Recently, layered transition metal dichalcogenide (TMD) superconductors approaching the two-dimensional limit have garnered significant attention for the enhanced phase fluctuations and electronic correlations. Investigating AMS in these systems, particularly in the absence of an external magnetic field, could offer valuable insights into the dimensionality-driven emergence of exotic quantum phenomena, including triplet Cooper pairing, phase fluctuation dynamics, and especially the recently discovered field-free superconducting diode effects. However, the field-free AMS has yet to be observed in TMD superconductors. Here, we report the dimensionality-tunable AMS near the superconducting quantum phase transitions in a layered TMD superconductor 2H-$Ta_2S_3Se$. In samples with thicknesses below 10 nm, we demonstrate magnetic field-driven AMS under external magnetic field, characterized by the vanishing of the Hall resistance and the presence of finite longitudinal resistance. Remarkably, an unexpected zero-field AMS emerges as the sample thickness is reduced to 3 nm. This AMS aligns well with the quantum vortex creep model and exhibits non-reciprocal transport behaviors, suggesting the onset of spontaneous time-reversal symmetry breaking accompanied by vortex motion as the system approaches the two-dimensional limit. Our findings open new avenues for exploring dimensionality-driven exotic superconducting quantum critical phases, and pave the way for a deeper understanding of zero-field superconducting diode effects.

___________________________________________

*These authors contributed equally to this work.
†Contact author: bincheng@njust.edu.cn
‡Contact author: miao@nju.edu.cn

Anomalous metallic state (AMS) has garnered significant attention in the research field of superconductivity [1-4], particularly in the contexts of the superconductor-to-metal transition and the superconductor-to-insulator transition induced by external stimuli such as magnetic field [5-10], structural design and electric doping, etc. The key features of AMS include the saturation of longitudinal resistance and the disappearance of Hall resistance at low temperature [2,11-14], both of which can be captured by two distinct theoretical frameworks. In the Bose metal model [15,16], such unusual electric transport behaviors are attributed to the collective mode of the phase glass. Meanwhile, in the quantum creep model [17-23], the flattened longitudinal resistance is interpreted as being due to the motion of quantum vortices. Notably, AMS typically emerges in superconductors with reduced dimensionality, where enhanced phase fluctuations and electronic correlation effects play significant roles [24-26]. To that end, the study of AMS with modulated dimensionality could give unique opportunities for investigating how superconducting fluctuation dynamics influence the dimensionality-tunable exotic phenomena such as triplet Cooper pairing [27-29], unconventional superconducting quantum criticality and the recently identified superconducting diode effects [7,9,30-40].

Due to the ease of controlling the sample thickness, layered van der Waals superconductors made from crystalline transition metal dichalcogenides (TMD), such as $NbSe_2$, $TaSe_2$, $TaS_2$ and $WTe_2$ provide an unprecedented platform for investigating the dimensionality-tunable AMS. Especially, when the dimensionality is reduced, the Ising nature [41,42], superconducting fluctuation as well as its dynamics [43,44], and electronic correlation effects could be significantly enhanced [45,46], leading to the emergence of various thickness-dependent quantum phenomena. This indicates that studying dimensionality-tuned AMS in TMD superconductors would offer new insights into these exotic quantum phenomena. However, the manifestation of AMS in these TMD superconductors typically requires the application of an external magnetic field, which limits the ability to study these dimensionality-driven exotic phenomena under field-free conditions.

In this work, we systematically investigate the dimensionality-driven behaviors of the AMS in TMD superconductor 2H-$Ta_2S_3Se$. With the disappearance of Hall resistance under an out-of-plane magnetic field, we confirm a magnetic field-induced AMS in samples with thicknesses approaching 10 nm. This AMS aligns with the quantum vortex creep model and exhibits non-reciprocal transport behaviors, which can smoothly transition to superconducting diode effects as the external magnetic field drops below the upper critical field. Remarkably, we discovered an AMS exhibiting field-free non-reciprocal transport behavior as the thickness was reduced to 3 nm. Our experimental results offer new insights into the unconventional superconducting quantum criticality and the field-free superconducting diode effects.

The $Ta_2S_3Se$ studied here is a layered superconductor with a 2H-type crystal structure (see Supplementary Note 1 and Fig. S1 [47] for the details about the synthesis and structural characterization). As shown in the inset of Fig. 1(a), a single layer of $Ta_2S_3Se$ consists of two layers of chalcogen atoms (S or Se) with a metal atom (Ta) layer sandwiched in between [48-49]. Notably, $Ta_2S_3Se$ exhibits higher superconducting critical temperatures compared to $TaS_2$ and $TaSe_2$ [43,50].

We totally fabricate eight $Ta_2S_3Se$ devices (labeled as devices #1 to #8) with varying thicknesses from 3.1 nm to 25.2 nm, with the optical images shown in Fig. S2 [47], to investigate the influence of dimensionality on superconductivity in $Ta_2S_3Se$. Here, the sample thickness is determined by the atomic force microscopy, as illustrated in Fig. S3 [47]. Then we measure the four-probe resistance $R$ under varying temperature, and plot the renormalize $R/R_n$ (where $R_n$ represents the normal-state resistance at 5 K) as a function of temperature shown in Fig. 1(a). For devices #1 and #2 with sample thicknesses above 10 nm, resistance sharply drops to zero at approximately 3.8 K. In contrast, for devices #3 to #8 with thicknesses below 10 nm, the temperature at which the resistance drops decreases with reduced thickness. We then extract the thickness dependent superconducting critical temperature $T_C$ (defined as the temperature at which the four-probe resistance drops to 1% of $R_n$), with the results shown in Fig. 1(b). Notably, the resistance does not completely vanish even below 2 K in devices #6 to #8 with sample thicknesses close to 3 nm, indicating the absence of the superconducting state even without external magnetic field. The critical magnetic field and critical current of $Ta_2S_3Se$ with different thicknesses are shown in Fig. S4 and Fig. S5 [47]. These observations of thickness dependence underscore the significant role of dimensionality in the superconducting behavior of $Ta_2S_3Se$ as the sample approaches the two-dimensional limit.

Next, we investigate the influence of sample thickness on the superconducting quantum criticality and corresponding anomalous metal phase by measuring the longitudinal resistance ($R_{xx}$) and Hall resistance ($R_{xy}$) as functions of magnetic field. In the thicker sample such as device #2 (14.5 nm), we observe the simultaneous emergence of finite $R_{xx}$ and $R_{xy}$ as the magnetic field is varied, as shown in upper panel of Fig. 2(a), indicating the absence of anomalous metal. In contrast, in device #3 with sample thickness of 9.4 nm [lower panel of Fig. 2(a)], $R_{xx}$ deviates from zero while $R_{xy}$ still remains zero, indicating the emergence of anomalous metal as the superconductivity is suppressed by external magnetic field. Such anomalous metal can be further confirmed by the saturation of $R_{xx}$ to a finite value and the vanishing of $R_{xy}$ with decreasing temperature under a fixed magnetic field, as shown in Fig. S6 [47].

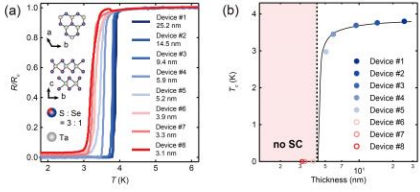

FIG. 1. (a) Normalized four-probe resistance versus temperature of Ta$_2$S$_3$Se samples with different sample thicknesses. (b) The superconducting properties of Ta$_2$S$_3$Se with different sample thicknesses.

To further investigate the AMS, we plot the temperature dependence of $R_{xx}$ and $R_{xy}$ for device #3 under various out-of-plane magnetic fields, following the Arrhenius convention, as shown in Fig. 2(b) and Fig. S7 [47]. Notably, as the temperature decreases beyond $T_{cross}$ {see Supplementary Note 2 for the details about the thermally activated flux flow (TAFF) [47]}, denoted by the black arrow in Fig. 2(b), the curves begin to deviate from the guide line representing thermally activated behavior. As the temperature further decreases, for the out-of-plane magnetic fields below 0.3 T, both $R_{xx}$ and $R_{xy}$ can drop to the minimal value within the measurement limit (below the orange dashed line), entering the superconductivity state. For the magnetic fields between 0.3 T and 0.7 T, only $R_{xy}$ {Fig. S7 [47]} reaches zero resistance within the measurement limit, indicating the presence of the AMS within this range of magnetic field.

To investigate the underlying mechanisms of the anomalous metal state, we further analyze the response of $R_{xx}$ to the magnetic field at 2 K. As shown in Fig. 2(c), we demonstrate that the magnetic field dependence of $R_{xx}$ follows the quantum creep model [19,21], which describes temperature-independent quantum tunneling of vortices (see Supplementary Note 3 for the quantum creep model and the Bose metal model [47]). Based on the observations from device #2, we obtain the phase diagram [Fig. 2(d)] to better illustrate the transition from the superconducting state to the normal state induced by the magnetic field. In this phase diagram, the upper boundary of superconducting state is determined by $T_c$ denoting the temperature $R_{xx}$ drops to 1% of its value at 5 K, and the lower boundary of the normal state region is determined by $T_{C2}$ (where $R_{xx}$ drops to 50% of its value at 5 K). In the transition region between these two states, the anomalous metal state and TAFF emerge. The boundary of the AMS, defined by $T_{AM}$, corresponds to the temperature at which $R_{xy}$ decreases to 1% of its value at 5 K. To summarize, the phase diagram indicates that in samples with thicknesses approaching 10 nm, the ground state remains superconducting. During the transition from the superconducting to the normal state, a magnetic field-induced AMS emerges, which aligns with findings in other systems such as NbSe$_2$, SnSe$_2$ and WS$_2$ [16,19,21].

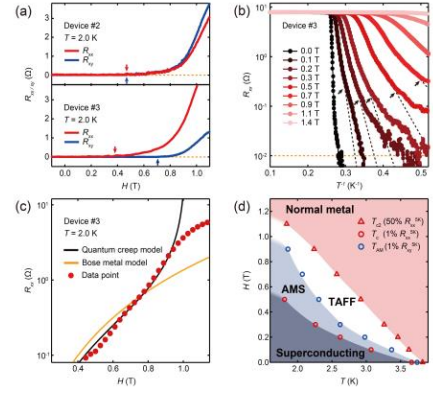

FIG. 2. (a) $R_{xx}$ or $R_{xy}$ as a function of magnetic field for device #2 (top panel) and device #3 (bottom panel) at 2 K. (b) Arrhenius plot for the temperature dependence of $R_{xx}$ measured at different magnetic fields. The orange dashed line indicates the minimum values within the measurement limit. (c) The magnetic field dependence of $R_{xx}$ in device #3, along with its fitting curves based on the quantum creep model (black solid line) and the Bose metal model (orange dashed line). (d) Magnetic field versus temperature phase diagram.

We then perform the transport measurements in device #8 with thickness of 3.1 nm to investigate the dimensionality-driven AMS approaching two-dimensional limit. As shown in Fig. 3(a), at zero magnetic field, $R_{xx}$ does not reach zero resistance but instead saturates at 1.95 Ω at low temperature, indicating the suppression of superconductivity in this thin sample. Meanwhile, as $R_{xx}$ exhibits a saturation trend, $R_{xy}$ vanishes, the anomalous metallic ground state under zero magnetic field is confirmed. This anomalous metallic ground state observed in Ta$_2$S$_3$Se at thicknesses approaching the two-dimensional limit is unique, which is absent in other layered TMD superconductors such as TaS$_2$, TaSe$_2$, and NbSe$_2$ [16,43,50]. The observation of this dimensionality-driven, zero-field anomalous metallic state opens unprecedented opportunities to explore exotic quantum states driven purely by dimensionality, quantum fluctuations and correlations, free from the influence of external magnetic fields.

We further check the magnetic field dependence of the anomalous metallic ground state by measuring $R_{xx}$ and $R_{xy}$ under varying magnetic field. The critical field can be determined by the temperature dependence of $R_{xx}$ and $R_{xy}$ under different out-of-plane magnetic fields, as depicted in Fig. S8 [47], analyzed following the Arrhenius convention. As shown in Fig. S9(e) [47], $R_{xy}$ starts to deviate from zero above 0.13 T, which can be defined as the critical field of the AMS ($H_{AM}$). When applied magnetic field is below $H_{AM}$, the $R_{xx}$ persists at a nonzero value and $R_{xy}$ disappears at low temperature, confirming the emergence of anomalous metal state. In contrast, when the applied magnetic field exceeds $H_{AM}$, both $R_{xx}$ and $R_{xy}$ no longer display the characteristics of the anomalous metal state. Moreover, we analyze the response of $R_{xx}$ to the magnetic field by fitting the quantum creep model in device #8, as shown in Fig. 3(b). To focus on the anomalous metallic state, the fitting was performed within the 0–0.13 T range. In this regime, the data align more closely with the quantum creep model than with the Bose metal model [47], suggesting that the underlying microscopic mechanism of the AMS observed in samples with different thicknesses is consistent, regardless of the presence or absence of a magnetic field.

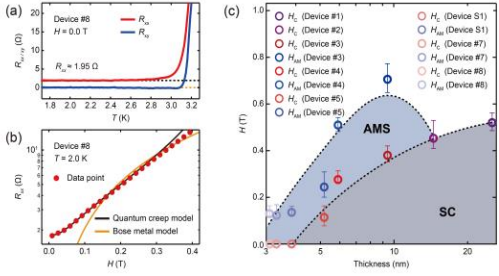

FIG. 3. The dimensionality-driven anomalous metallic state. (a) $R_{xx}$ or $R_{xy}$ as a function of temperature at zero magnetic field for device #8. (b) The magnetic field dependence of $R_{xx}$ in device #8, along with its fitting curves based on the quantum creep model (black solid line) and the Bose metal model (orange dashed line). (c) Phase diagram of thickness versus magnetic field.

We then summarize the dimensionality-driven AMS by plotting the phase diagram of magnetic field versus sample thickness, as shown in Fig. 3(c). Here, the critical fields $H_C$ and $H_{AM}$ are extracted from the curves of $R_{xx}$ and $R_{xy}$ versus magnetic field in devices #1-5, #7-8 and S1 {Fig. 2(a) and Fig. S9 [47]}. The trend line defined by $H_C$ marks the upper boundary of the superconducting phase, while the trend line determined by $H_{AM}$ delineates the upper boundary of the anomalous metallic state. We observe that the superconducting ground state is rapidly suppressed as the dimensionality decreases, eventually disappearing. Meanwhile, the AMS emerges under the application of a magnetic field at a thickness near 10 nm, and transitions to a zero-field ground state at a thickness of approximately 3 nm. Two mechanisms—quantum phase fluctuations [25] and formation of inhomogeneous superconducting states [14] due to electronic phase separation [8, 19]—may both contribute to the development of the zero-field AMS as the dimensionality is reduced. Further insights into its microscopic nature may be obtained through local probe techniques such as scanning tunneling microscopy and SQUID-on-tip, which are promising tools for future investigations. The presence of AMS at zero magnetic field gives unprecedent opportunities for investigating the underlying mechanism of exotic superconducting phenomena with spontaneous TRS-breaking such as zero-field superconducting diode effect.

Finally, we measure the voltage versus current ($V$-$I$) curve and second harmonic signals in devices #5 (5.2 nm) and #6 (3.9 nm), which exhibit a superconducting state and an AMS at low temperature, respectively, to investigate their nonreciprocal transport properties. As shown in Fig. 4(a), we first perform current sweeps at 2 K under zero magnetic field in device #5, which is fully in the superconducting state, and monitor the longitudinal voltage signal. The voltage versus current curves for forward and reverse sweeps overlap, indicating the absence of a superconducting diode effect. By applying a small out-of-plane magnetic field to break TRS, the superconducting state is not yet destroyed and the critical current [Fig. 4(b)] exhibits nonreciprocity, like the magnetic-field-assisted superconducting diode effect observed in $NbSe_2$ [32]. Furthermore, as shown in Fig. 4(c) and Fig. 4(d), with increasing applied magnetic field, the superconducting ground state transitions into the AMS, where a noticeable inequivalence between the positive ($I_s^+$) and negative ($I_s^-$) bias regimes appear, which indicates the asymmetric pinning effect may also influence the quantum vortex motion in AMS and leads to nonreciprocal transport [51]. Notably, in addition to applying a magnetic field, increasing the temperature can also drive the device #5 into the AMS. Dramatically, although the TRS is not explicitly broken since external magnetic field is absent, we still observe nonreciprocal transport behaviors in this AMS under zero magnetic field, as confirmed by the inequivalence in the $V$-$I$ curves shown in Fig. 4(e).

The evolution of nonreciprocal transport behaviors in both the superconducting state and AMS, under varying magnetic fields and temperatures, can be further confirmed by measuring the corresponding changes in the second harmonic signal of longitudinal resistance. As shown in Fig. 4(f), the second harmonic signal appears and extends across a magnetic field range at varying temperatures. Specifically, at temperatures below 2.8 K, the second harmonic signal vanishes near zero magnetic field, while the field range exhibiting a pronounced second harmonic response closely corresponds to the AMS. These second harmonic signals likely arise from asymmetric dissipation caused by current-driven vortex motion, which is linked to vortex-related mechanisms underlying the formation of the AMS [52]. Meanwhile, at 3.2 K and 3.4 K, a prominent second harmonic signal appears at zero magnetic field, which is consistent with the inequivalence in the *V-I* curve at 3.2 K. This temperature dependence, under fixed extrinsic conditions, suggests that the nonreciprocal signal is intrinsic to the AMS phase rather than caused by static sample inhomogeneity, and points to the possible emergence of dynamic orders from spontaneous TRS-breaking in this thermally activated AMS.

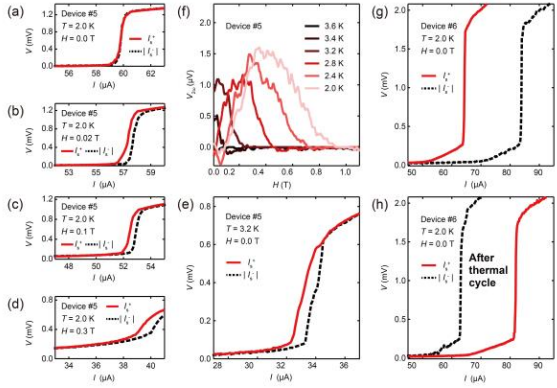

FIG. 4. Nonreciprocal transport behavior of the anomalous metallic state. (a-d) *V-I* curves at 2 K for device #5 under magnetic fields of 0.0 T (a), 0.02 T (b), 0.1 T (c) and 0.3 T (d). (e) *V-I* curves at 3.2 K for device #5 under zero magnetic field. (f) Second harmonic signal as a function of magnetic field at different temperatures for device #5. (g) *V-I* curve at 2 K for device #6 under zero magnetic field. (h) *V-I* curve for device #6 with reversed nonreciprocal polarity after thermal cycle under zero magnetic field.

The observation of field-free nonreciprocal transport behavior in the thermally activated AMS motivates us to investigate the nonreciprocity in the thin sample (device #6, with a thickness of 3.9 nm), which exhibits a field-free AMS rather than a superconducting state at low temperatures. As shown in Fig. 4(g), at 2 K and zero magnetic field, device #6 remains in the AMS, and its voltage-current characteristics clearly exhibit pronounced inequivalence, indicative of field-free nonreciprocal transport behavior, as shown in Fig. 4(g). This observation of field-free nonreciprocity in device #6 can be further confirmed by the prominent second harmonic signal near zero magnetic field (Fig. S10 [47]). Similar nonreciprocal transport behavior was also observed in two additional devices {Fig. S10(b–d) [47]}, highlighting the consistency of this phenomenon across multiple samples. These observations suggest that the current-driven quantum vortex motion in AMS can be asymmetric, which could result from spontaneous TRS-breaking, even in the absence of an external magnetic field. Note that such behavior has not been reported in other TMD superconductors and thus provides a new platform for exploring the microscopic origin of zero-field nonreciprocal transport, especially superconducting diode effect, in layered quantum materials.

Moreover, we observe that the polarity of the nonreciprocal transport behavior in the *V-I* characteristics shown in Fig. 4(g) is positive, meaning that the voltage during the forward scan exceeds that of the reverse scan. Notably, this polarity can be reversed through a thermal cycling process. After heating to 300 K and cooling back to 2 K, the polarity may either switch or remain unchanged. Fig. 4(h) presents the voltage-current curve, where the polarity reverses to negative after a thermal cycle, excluding the influence of any residual environmental field and thus supporting the prominent role played by the spontaneous TRS-breaking. Meanwhile, this zero-field nonreciprocal transport behavior cannot be switched by external magnetic field. This result suggests the existence of dynamic orders in an internal TRS-breaking background, reminiscent of the zero-field superconducting diode effect observed in cuprate superconductor [40], except that the ground state here is an anomalous metal rather than a superconductor. In addition, other scenarios, including a pair-density-wave phase [36,53], chiral charge ordering [54], and boundary current [37] may also contribute to this TRS-breaking, which requires more theoretical investigations to elucidate in the future. Taken together, these findings underscore that the microscopic mechanism behind the non-reciprocal transport in $Ta_2S_3Se$ remains an open question requiring further in-depth study. Thus, these nonreciprocal transport behaviors observed in the AMS, particularly in the zero-field condition, could give some clues for understanding the field-free superconducting diode effect in layered superconductors.

In conclusion, we present a dimensionality-tunable AMS in a layered Ising superconductor 2H-$Ta_2S_3Se$. The evolution of the AMS under varying magnetic fields follows the quantum creep model describing vortex motion. Remarkably, as the sample approaches the two-dimensional limit, the anomalous metallic state can persist even without an external magnetic field, a behavior not observed in other TMD superconductors such as $NbSe_2$, $TaS_2$, or $TaSe_2$. This AMS dramatically exhibit zero-field non-reciprocal transport behaviors, suggesting the emergence of spontaneous TRS-breaking. Our findings open unprecedented opportunities to explore exotic quantum states near superconducting quantum criticality, characterized by enhanced quantum fluctuation dynamics and unconventional mechanisms of spontaneous symmetry breaking.

*Acknowledgments*—This work was supported in part by the National Key R&D Program of China under Grant 2023YFF1203600, the National Natural Science Foundation of China (62122036, 12322407, 62034004, 61921005, 12074176), the National Key R&D Program of China under Grant 2023YFF0718400, the Leading-edge Technology Program of Jiangsu Natural Science Foundation (BK20232004), the Strategic Priority Research Program of the Chinese Academy of Sciences (XDB44000000), the Fundamental Research Funds for the Central Universities (14380227, 14380240, 14380242), and the Jiangsu Funding Program for Excellent Postdoctoral Talent (2022ZB65). F. Miao and S. -J. Liang would like to acknowledge support from the AIQ Foundation and the e-Science Center of Collaborative Innovation Center of Advanced Microstructures. The microfabrication center of the National Laboratory of Solid State Microstructures (NLSSM) is also acknowledged for their technical support. We thank Yan Luo and Chao Zhu for technical support in material structure characterization. We also acknowledge Wei Wei and Fanqiang Chen for assisting with the RC filter technique.